# Multiple Conceptual Coherences in the Speed Tutorial: Micro-processes of Local Stability

Brian W. Frank, University of Maine, 5709 Bennett Hall, Orono, ME 04469-5709, brian.frank@umit.maine.edu

**Abstract:** Researchers working within knowledge-in-pieces traditions have often employed observational approaches to investigate micro-processes of learning. There is growing evidence from this line of work that students' intuitive thinking about physical phenomena is characterized more so by its diversity and flexibility than its uniformity and robustness. This characterization implies that much of the dynamics of students' thinking over short timescales involve processes that stabilize local patterns of thinking, later destabilize them, and allow other patterns to form. This kind of "change" may only involve dynamics by which the system of intuitive knowledge settles into various states without changing the system structure itself. I describe a case study in which a group of college students shift their thinking about motion several times during a collaborative learning activity. Instead of focusing on micro-processes of change, I describe these dynamics in terms of mechanisms that contribute to local stability of students' conceptual coherences.

## Introduction

Conceptual change research in the seventies and eighties largely began focused on characterizing the singular and often flawed ideas that children and students espouse about domain-specific phenomena (Driver & Easley, 1978; Posner, Strike, Hewson, & Gertzog, 1982; Carey, 1986). From this perspective, it was largely assumed that unitary knowledge structures–naïve theories (McCloskey, 1983), alternative frameworks (Clement, 1983), or ontological categories (Chi, 1992)–describe important aspects of novice misunderstandings and that these misunderstanding must give way to other structures in a progression toward expertise. In contrast, more research in the past two decades has focused on characterizing the *diversity* of ideas that children and students employ for making sense of physical phenomena and the nuanced *contextuality* of these varied ideas across time and setting (Strike & Posner, 1992; Smith, diSessa, & Roschelle, 1993; Hammer, Elby, Scherr, & Redish, 2004). From this perspective, it has largely been assumed that a substantial component of expertise involves the gradual refinement and repurposing of existing knowledge–phenomenological primitives (diSessa, 1993), conceptual and epistemological resources (Hammer, 1996), or even pieces of ontological knowledge (Gupta, Hammer, and Redish, accepted).

Both of these perspectives have largely made *change* a central problem to be explained by research. Much of the prior research has focused on broad processes and conditions for change, while the latter has focused on micro-processes of change. In this paper, I focus on describing student understandings as reflecting *multiple local coherences* (Hammer et al, 2004; Rosenberg, Hammer, & Phelan, 2006; Scherr & Hammer, 2008). This specific term is intended to capture the notion that understanding and behavior are often quite variable, while still exhibiting stabilities that are local to moments and settings. While much of this existing work has focused on epistemological coherences, I focus here on describing local coherences that concern the substance of students' thinking. This perspective shifts attention away from seeking only explanations of change and toward explanations of how understandings cohere in specific moments and settings. Locally coherent patterns of student thinking are to be explained in terms of mechanisms that stabilize these patterns. This perspective is explored through an analysis of students' multiple conceptual understandings during a collaborative learning activity in an introductory college physics classroom.

## Interpreting What Distance Means in the Speed Tutorial

Before discussing the theoretical and epistemological considerations for this work, I want to motivate this discussion through an example of student reasoning that is the focus of this study. Two vignettes below are taken from a student discussion during a collaborative learning activity in introductory college physics. The students are in a tutorial—a common instructional format in introductory physics classrooms where students are expected to work together through a conceptual worksheet. In this particular tutorial (a modified version of a tutorial from McDermott, Shaffer, and the Physics Education Group at the University of Washington, 1998), a group of four college students discuss "tickertape" representations of motion.

Tickertape refers to a long strip of paper that can be used to create a record of an object's motion by attaching the paper to the moving object. They are commonly used in physics classrooms to teach students concepts about velocity and acceleration. The record of the motion is produced as the paper is pulled through a tapping device that marks the paper at a constant rate. In this tutorial, strips of tickertape represent the motion of a cart that was recorded prior to class (see Figure 1 for a cartoon schematic). Each student has been given a small strip cut from the entire tickertape that represents the same amount of time (i.e., has the same number of dots) but represents different speeds (i.e., have varying length).

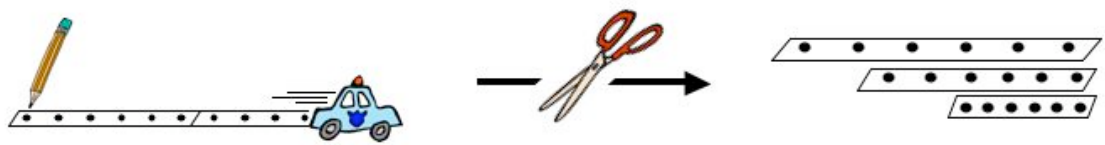

Figure 1: Cartoon Schematic for How Tickertape Strips were Made Prior to Class

Below, the students discuss how the amount of time taken to generate each of their different strips compares, and incorrectly decide (as many groups initially do) that the shorter strips represent less time.

> Beth: Obviously, it takes less time to generate the more closely spaced dots.
> John: So you are saying it takes less time to make the shorter segments?
> Beth: Yeah.
> Kate: How can you tell?
> Beth: You can tell because it's a shorter distance.
> Paul: It's a shorter segment.

The students discuss this conclusion for several minutes, pointing out other features of the strips that support this conclusion. They soon move on to the next part of the worksheet that involves doing some calculations based on distance measurements. After finishing these calculations, the students are prompted by a specific question in their tutorial workbook to reflect upon their assumptions in doing these calculations.

> Kate: I think we also assumed in that these were made by the speed at which the paper traveled through the tapper, which was different for each paper.
> John: Right, cause if you move it really fast then *[quick movement of hand]*
> Beth: That's true! It could depend on how fast the ribbon was pulled *[moves hand]*
> Kate: We're assuming that umm-
> Beth: That the length is proportional to-
> Kate: -The speed at which the ribbon was pulled through.

These two vignettes involve student understandings that are substantively quite different. In the first, students interpret distance as conveying information about time. In the second, students interpret distance to mean speed.

It seems natural, given the obvious change in students' interpretation of what distance means, to ask, "What caused the students to change their thinking?" This focus seems especially relevant since they changed their understanding from an incorrect one (thinking that the distance indicates time) to a correct one (that the distance indicates speed). We might consider several plausible accounts of what caused the change: (i) the particular question from the worksheet triggered them to think meta-cognitively about what they had been doing, (ii) engaging in the measurement exercise enabled the students to attend to and coordinate new aspects, and (iii) Kate was privately thinking the correctly the entire time and only now found a way to have her ideas introduced and considered. Each of these answers may hold some truth in explaining the change in students' collective understanding. The question about change, however, may be misleading.

It turns out that the students don't change their thinking on the matter this one time. Several minutes after the end of the above transcript, Beth brings up again the idea that the shorter strips take less time. Kate, who initially introduced the idea that the speed causes the different distances, seems to now agree with Beth. As the tutorial goes on, the group works to regenerate their understanding that distance must be related to the speed and not the time. However, this is not the last change. As the students go back to a previous page to erase their wrong answers, one of the students is again convinced of their earlier idea. It's not entirely clear that the students ever resolve the issue completely during the instructional period. Instead, their understanding seems to vary between these two distinct ways of making sense of the strips of paper as representations of motion.

For the researcher, trying to understand the cause of just a single moment of change in the students' thinking now becomes a problem of explaining multiple changes back and forth. The seemingly straightforward question, "What caused the students to change their thinking?" may not only be difficult to answer, it may be misleading in its premise. For starters, there might not be a simple or single causal explanation for any of these transitions. In addition, the apparent change in the students' thinking might not reflect change to the structure of students' intuitive knowledge at all. Instead, the changes we observe might only reflect dynamic transitions among competing understandings that each exhibits some local stability. Understanding the local stabilities themselves becomes a central problem for researchers to address. The aim of this paper is to describe how the students' initial understanding may be stabilized through a variety of different mechanisms.

## Theoretical Orientations

The shift in perspective I am advocating coincides with an underlying assumptions that distinct patterns in student thinking reflect dynamic stabilities and that these stabilities require explanation. The stability of student thinking need not be due to the existence of a robust belief or mental category, but rather due to how real-time activity creates self-sustaining patterns of thought. The notion of real-time activity has often been discussed dichotomously with respect to human behavior and learning—either encompassing individual, cognitive activity (e.g., connectionism) or as distributed, social activity (e.g., activity theory). In this section, I briefly discuss relevant aspects of these individual and activity perspectives.

### Activity among Knowledge Agents

From individual cognitive perspectives, complex knowledge systems approaches describe activity as occurring among fine-grained knowledge structures. Minsky (1986), for example, conceptualized the individual mind as a society of mindless "agents"—numerous in kind and in their interactions. Agents, like schemata more generally conceived (Bartlett, 1932/1995; Rumelhardt, 1980), can be thought of as cognitive elements for perceiving, doing, and remembering. These pieces come together in various ways to generate and represent thinking. One kind of knowledge piece that has been discussed quite extensively is diSessa's (1993) phenomenological primitives—bits of knowledge reflecting our intuitive sense of mechanism. In the above transcript, we might think of the students' inference about the shorter strips taking less time as resulting from the intuition that *less distance implies less time*. This bit of knowledge for relating space and time is part of how the students interpret and make sense of the strips as representations of motion.

Beyond just describing kinds of knowledge, diSessa distinguishes between two properties of intuitive knowledge systems that he calls *cuing priorities* and *reliability priorities*. These two properties drive mechanisms that activate and stabilize ideas in particular contexts. Phenomenologically, cuing priorities describe the likelihood that a given piece of knowledge will activate in a given context, while reliability priorities describe the likelihood that a given piece will remain active in a given context. In the above example, we see the idea "shorter strips take less time" becoming focal in the conversation and then persisting for several minutes beyond the transcript.

Structurally, if one thinks of knowledge elements in terms of a network, reliability is related to the number of activation pathways leading away from and then back to a given knowledge element. In other words, a given idea can be made more stably active (in a given setting) because other "nearby" ideas support its sustained activation. Thus, the relationships and interactions among many ideas serve as one kind of mechanisms by which patterns of thinking are self-sustaining over a period of time. With the transcripts above, we whether other ideas arise during their conversation that help to support their continued thinking that the shorter strips take less time, as well as how the context supports the cuing of those ideas.

### Activity among Participatory Agents

From a more distributed and social perspective, activity-theoretical and other situated approaches describe activity as occurring among persons and artifacts, not among entities of the mind. In many respects, this view represents an oppositional stance to the above knowledge-based accounts. Lave (1988), for example, in describing the variety of mathematical practices, denies the existence of mental constructs such as arithmetic. Despite this dichotomy, there are those who do attempt to capture both knowledge and participation views in their accounts of students' thinking (e.g., Roschelle, 1998) or have advocated for their commensurability.

One major difference between knowledge- and participation-based accounts concerns the typical unit of analysis. The unit of analysis for situated perspectives is often described as the entirety of *persons-in-settings* (rather than individuals or individuals' knowledge). It is more focused on observable unfolding social activities (than knowledge use) and how these activities arise within social settings that are defined by the activity (and not by platonic outsider descriptions). Beyond just focusing on humans, these perspectives focus on describing how material artifacts mediate aspects of that activity (Engestrom, 1987), provide affordances for action (Greeno, 1998), and shape semiotic fields (Goodwin, 2000). In the vignettes above, the students do not only have different understanding. They also interact with each other and their settings in quite different ways.

Different patterns of action and social collaboration, taking place within changing material settings, can be understood to constrain and support activities that also differ in terms of the *substance of the ideas* to which activity is oriented. Embodied cognitive perspectives, for example, emphasize how human understandings have a strong basis in bodily action (e.g., Thelen & Smith, 1994). Pointing is an action that might support ideas for understanding objects and their spatial relations, while dynamically moving one's hand through space might better support ideas for understanding trajectory and cause and effect relations. Similarly for social collaboration, establishing joint attention as part of collaborative action might afford different epistemic activities than establishing mutual attention. Such social cues help individuals to *frame* (Scherr & Hammer, 2009) activities.

Taken together, the affordances of artifacts and the patterns of participation that take shape around them serve as mechanisms by which activities stabilize in settings. By carefully considering the substance of ideas to which these activities orient as well, the analytical task undertaken in the next section is to explain the stability of students' understanding in terms of both knowledge- and participation-based mechanisms.

## Explaining the Local Stability of Reasoning

The goal of this section is to develop an explanatory account of the stability of the students' initial thinking in terms of real-time processes. I describe several knowledge- and participation-based mechanisms that plausibly contribute to the stability of students' thinking and connect these claims to specific evidence.

## Knowledge Based Mechanisms

### Contextual Feedback from Setting

One way in which the students' initial thinking that "shorter strips take less time" stabilizes is due to how particular micro-features of the context support the continued activation of underlying intuitive knowledge. Several aspects of the particular context in which students discuss their ideas support the persistence of the idea that *less distance implies less time*: the specific language of the worksheet question, the salience of length differences among the tickertape strips, and a congruence of part-whole relationships on the strips.

The particular question from the worksheet reads, "How does the time taken to generate one of the short segments compare to the time to generate one of the long ones?" Beth immediately states, "Obviously, it takes less time to generate the more closely spaced dots." The particular phrasing of the question not only draws attention to the distance features of the strips, it uses the words "shorter" and "longer" to do so. These specific words are relevant, because they are used flexibly in everyday language to both refer to distance and time (but no other concepts). The ambiguity of these words closely ties with the intuition that *less distance* ("shorter") *implies less time* ("shorter"). The students repeatedly use those words while discussing their initial ideas.

The students' understanding is also supported by features of the tickertape strips. By far, the most salient feature of the strips is that they are different lengths. One can observe this when standing far away, without closely inspecting the details of where the dots are located. The salience of this feature supports students' initial and sustained attention to distance features of the strips, which has been shown to preferentially cue the idea that *less distance implies less time* over other competing ideas when students are asked to made judgments about duration (Frank, Kanim, Gomez, 2009). In the transcript, after Beth initially gives her answer, the students explain that they can tell its less time because the strips are shorter. As they explain this, the students can be seen reaching toward the strips and indicating distances on the strips with their fingers and pencils. These statements and gestures show that students are closely attending to this salient feature in support of their initial understanding.

A second feature of the strips that contributes some stability arises from the fact that each of the strips has the same number of dots. When inspecting the strips, students do not only attend to the total length of the strips, but also to the length of the spaces between dots. Because each strip has the same dots, this means that the longer strips are made up of longer pieces, and the shorter strips are made up of shorter pieces. As students shift their attention from the entire length to the length of the parts (or vice versa), students see the same information. Shorter strips or shorter parts imply a shorter amount of time. Evidence that this shift in attention does not disrupt the students' thinking, and may in fact stabilize it, comes from a part of their conversation that happens shortly after the end of the first transcript above.

Kate: When we are talking about segments, are we like not thinking about how long the total paper is? Are we just looking at the marks. Are we supposed to be considering
John: I'm guessing they like mean from here to here. *[pointing with pencil between marks]*
Kate: Like I wonder why like the papers are all different lengths.
Beth: Cause none of these papers are the exact same si-ize. Except for these two *[pointing]*
Paul: Right because I think *[moving hand toward the center]* they all have the same amount of dots *[pointing to several locations on one of the strips]*.
Beth: Oh-oh
Paul: I think they all have six dots.
Beth: Oh do, they?
Kate: Is that true? 1, 2, 3, 4, 5, 6 *[pointing to successive dots on a strip]*
Paul: So, it's a shorter amount of time for a shorter piece of paper

Kate asks if they are supposed to be looking at the distance between the "marks" or the total length. As the students coordinate between these two features (realizing that there are six dots on each), the students maintain their understanding that the shorter strips take less time.

### Structural Feedback from Other Knowledge

The students' initial understanding is also stabilized through the recruitment of other ideas that support its persistence. Two specific ideas arise together with the idea that the shorter strips take less time. One is that *bunched up means faster,* and the other is *faster implies less time*.

Students attend to the length of the entire strip and to the length of its parts, but they also attend to the "density' or overall proximity of the dots to each other. These students, as well as many others, describe the bunched up patterns as "faster" and the more spread out patterns as "slower". The words "faster" and "slower" used here is another example of ambiguous and flexible use word meaning, similar to how "longer" and "shorter" are used in reference to the distance and time. The students notice "more dots happening in less space" and describe this pattern as "faster". However, the same word is used by these students to describe a greater speed and also a lesser the amount of time, and therefore provides some ambiguity about its meaning.

Both immediately before and after the students discuss that the shorter strips takes less time, several of the students refer to the shorter strips (with the more closely spaced dots) as faster.

> Beth: Nobody has the exact same rate
> John: *[pointing to the shortest strip]* So I guess that's the fastest.
> … (and then later in their discussion)…
> Beth: How do you know how to arrange –
> Kate: Ahh. The shorter the segments the faster the speed
> John: Yeah.
> Paul: Also shorter the paper, it's the same thing.

Here we see the students using the word faster to identify shorter segments. The idea *bunched up means fast* can be understood to work together with another idea that arises. In this case study, and many others described elsewhere (Frank, 2009), we see evidence that students discuss the idea that *faster implies less time*. Beth explicitly discusses this idea during their discussion at a slightly later time when she says, "Wouldn't yours be going slower than mine, because it took more time to make that same?" Her statement implies that a shorter strip goes slower because it takes less time, which she and the other students had established earlier as being known because the strip was shorter (bunched up). A single strip is thus conceived of as being short, bunched, and fast, with the words "short" and "fast" having multiple meanings that overlap.

These three elements of intuitive knowledge–*less distance implies less time, bunched up means faster*, and *faster implies less time*– are all connected through their linguistic and conceptual overlap. Ideas of *less distance* and *less time* are bound by a linguistic overlap with the words "longer" and "shorter". *Less distance* and *bunched* are bound by their conceptual overlap with a spatial sense of proximity. *Bunched* and *less time* (and even notions of speed) are bound by their linguistic overlap with the word "faster" and "slower". This network of ideas can be understood to exhibit stability through the mutual relationships among its parts.

These knowledge-based mechanisms describe how relationships among specific elements of knowledge and contextual features provide some stability to their understanding, contributing to its persistence over time. In the following section, I describe aspects of activity that provide stability in different ways.

## Participation Based Mechanisms

The previous section largely focused on mechanisms that concern how knowledge elements remains active due to interactions with features of context and with other elements of knowledge. In this section, I discuss how the students' thinking is stabilized through processes by which students interact with each other and with artifacts.

### Feedback among Interactional Behaviors

Students' collective interactions with each other and with various artifacts around them show distinct patterns that are relevant to understanding their activity. For the first fifteen minutes, the students collectively move in and out two broad clusters of interactive behavior. These behaviors largely occur together among all the individuals, with some exceptions. The two distinct patterns are (i) behaviors oriented toward their worksheets (see Figure 2)- a pattern described by Scherr and Hammer (2009)- and (ii) behaviors oriented toward the strips at the center of the table (see Figure 3).

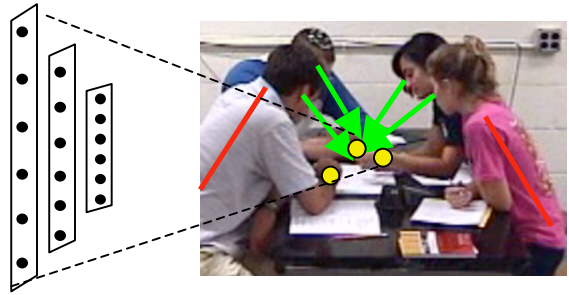

Figure 2. Student Behaviors Oriented toward the Strips

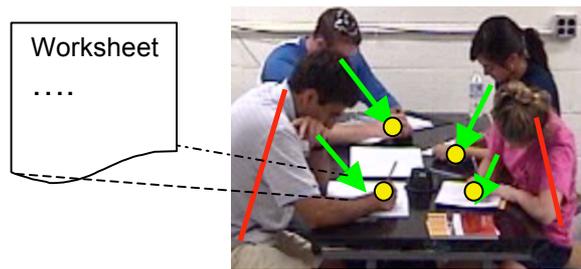

Figure 3. Student Behavior Oriented toward Worksheets

These behaviors support important aspects of the students' activity, which establish and stabilize a setting for their discussion. The tickertape-oriented behaviors are characterized by patterns of interaction that involve directing gaze, gesture, and bodies toward the strips at the center of the table. Students lean in, look inward, and position their hands at the center of the table. The students' clustering in this behavior is dynamically stable in the sense that any one student's behavior is coupled to the behaviors of the others. As students lean toward the center and point to strips and locations, other students look in that direction, and mirror these actions by leaning in and pointing to features as well. While individual students opt in and out, the collective structure of participation persists over much longer time scales. The worksheet-oriented behaviors are characterized by different patterns of interaction: hunching over, looking down, and using hands to write and point to worksheet locations. This pattern also exhibits a dynamic stability.

What is striking about the students' behavior during this time is the remarkable synchrony with which they move in and out of these two patterns. It is a highly coordinated (but not centrally directed) activity of noticing and describing patterns and then writing down answers to specific questions. This activity of establishing joint attention to community objects may be understood to couple to the substance of their ideas as well. Their initial understanding during this time concerns inferences drawn from noticing and comparing features of the strips (e.g., *short distance means short times* and *bunched up dots means fast*). These externalized behaviors for establishing joint attention are part of the dynamic by which the students access ideas and employ strategies used for noticing, describing, and comparing patterns across the strips.

Importantly, during this time, the students' neither discuss aspects of the physical motion of the strips, nor do they engage in any significant moments of mutual attention, as they do later. The above behaviors contrast with their later behaviors, where they orient their behaviors to each other—sitting up right, looking to each other, and gesturing in animated ways.

## Material Affordances and Constraints

The two patterns of collective behaviors above, which help to characterize the students' initial activity, take place in a particular material setting that provides affordances for these behaviors. The students themselves play a role in constructing this setting. In particular, during the time that the students' initially persist in thinking that the shorter strips take less time, the strips are located and arranged at the center of the table. The students moved the strips to this location and arranged them in order by length at the very onset of the tutorial. This centralized location allows the students to collectively orient to the strips and sustain an activity of pointing out and describing features of the strips.

The arrangement in the side-by-side fashion serves two purposes. First, it offloads much of the task of noticing and remembering to the setting. Second, the arrangement has affordances for particular kinds of gesture. Students gesture in various ways (pointing with a finger, indicating a space with two fingers, etc), but all of the gestures are deictic. These embodied actions may support students' ideas for thinking about objects and spatial relations among parts (the density of dot, the number of dots, etc.)

The anchoring of their activity to the strips as material artifacts and the dynamic stability of their collective behaviors feed into many of the knowledge-based mechanisms described in the previous section. Their individual behaviors couple to each other and to material objects. These behaviors help to sustain particular patterns of attention and action upon the strips. These behaviors support their use of specific intuitive knowledge. In one sense, we can think of the stability of students' thinking as coupled to a physical stability of the world. The strips, once placed at the center of the table, remain put until some actively moves them again. Similarly, the students have located their worksheets so that they are closely hugged to their bodies, making it easy to coordinate their dual activity between noticing and describing, and filling out their worksheets. The students seamlessly move back and forth between these two patterns of behavior repeatedly during the first ten minutes of tutorial. Their careful attention to just the visual properties of the strips (i.e., not to any sense of motion) in order to answer specific worksheet questions takes place in a dynamic setting that supports its persistence.

Just as students' thinking and behavior during the first episode contrasted with their later thinking and behaviors, the students' material setting later is different as well. Their later understanding happens in a setting where the students have separated and decentralize the strips. This change to their material settings provides new opportunities for embodied action and participation that did not occur before. In the second vignette, where Kate explains how the strips were made, Kate can be seen lifting a strip off the table and physically enacting the pulling of the strips by the cart. In doing so, the other students in the group look to Kate. The students engage in a new pattern of mutual attention as they discuss aspects of the physical motion which made the strips. This is the first time they engage in significant mutual attention, and is part of the dynamic by which they in articulate and listen to more complex ideas.

## Discussion

The primary goal of this paper is to characterize the stability of a single group of students' initial thinking during a collaborative learning activity in terms of a variety of micro-processes that together help to sustain one possible understanding for several minutes. A full analysis of these dynamics would also need to include an account of the mechanisms that contribute to their later thinking as well, but is beyond the scope of this paper. Given the dynamic changes apparent in their understanding during the tutorial, I claim that it is productive to focus on explaining the local persistence of understanding, rather than to begin with focus on causes of change.

One distinguishing feature of this analysis is its focus on explaining this stability both in terms of how elements of students' intuitive knowledge are reliably activated in the setting and in terms of how students' interactional behaviors dynamically constitute and stabilize aspects of this setting. I claim that the students make their initial inference about the tickertapes based on a shared intuition that *less distance implies less time*. There is evidence that each of the students makes sense of the situation using this idea at various times, and often times they do so together. I describe how particular features of the context (as characterized by the existence of specific features on worksheet question and on the strips themselves) contribute to activation of that knowledge, and also describe how other knowledge recruited to make sense of the situation provide a stable "network" of ideas for making sense of the patterns they notice. This network of ideas, however, arises within a social context of discussing patterns and writing down answers to worksheet questions. During their initial discussion, the students orient their gazes, gestures, and bodies toward the collection of the strips at the center of the table and toward the worksheets in front of them; but they do not engage in behaviors of mutual attention until much later. This activity of noticing and describing patterns to be written in their worksheets is both stabilized by the students' specific behaviors in this activity (to which they collectively orient) and by the location and arrangement of material artifacts that are central to the activity.

This characterization, while focused on explaining the stability of the students' initial pattern of thinking, does provide insight into the dynamics of their subsequent understanding as well. Instead of seeking explanations in terms of "causes" for the change, we can, instead, see how the various constraints imposed by their activity change. These changes in the constraints of their activity provide opportunities for new patterns to take hold with their own local stability. The students' new thinking co-evolves with changes to their patterns of attention (now to physical mechanisms involved in the motion), changes to the location of material objects (the strips become separated and within several of the students hands off the table), and to the patterns of interactional behavior (now they look at each other as they articulate complex ideas). Just as there may be not single cause for the local stability of the students' thinking, their may be not single explanation for why the students' thinking changes either. Across the entire tutorial, there are many changes that are more-or-less continuously changing over time. Coming to understand which aspects of activity are merely ephemeral and which continue to exert influences is a major goal of further research in this direction.

Beyond this particular case, considering students' intuitive thinking as reflecting multiple conceptual coherences may be useful for describing stabilities of thinking and behavior that seem to persist on longer time scales (e.g., Why do students appear to hold on to robust misconceptions?). It may be possible to both account for the variability of student understandings while also describing how, in particular contexts, it can settle into patterns indicative of common and robust misconceptions. This program might be achieved through careful attention to the multitude of cognitive and social mechanisms that contribute to the local stability of reasoning as it occurs within activity.

## Acknowledgments

The research described here was conducted at the University of Maryland. I'd like to thank Rachel Scherr, David Hammer, Joe Redish, Andy Elby, Ayush Gupta, Luke Conlin, Renee Michelle Goertzen, and Michael Wittmann for their helpful conversations. The research has been funded in part by the National Science Foundation under Grant Nos. REC-0440113 and REC-0633951. Any opinions, conclusions, or recommendations expressed in this material are those of the author and do not necessarily reflect the views of the National Science Foundation.